\documentclass[10pt]{article}
\usepackage{latexsym}
\usepackage{graphicx}
\usepackage{times}
\begin{document}
\title{Einstein's concept of a clock and clock paradox}
\author{Wang Guowen\\College of Physics, Peking University,
Beijing, China}
\date{January 25, 2005}
\maketitle
\begin{abstract}
A geometric illustration of the Lorentz transformations is given.
According to similarity between space and time and correspondence
between a ruler and a clock, like the division number in a moving
ruler, the tick number of a moving clock is independent of its
relative speed and hence invariant under the Lorentz
transformations. So the hand of the moving clock never runs slow
but the time interval between its two consecutive ticks contracts.
Thus it is Einstein's concept of slowing of the hands of moving
clocks to create the clock paradox or twin paradox. Regrettably,
the concept of the clock that Einstein retained is equivalent to
Newton's concept of absolute time that he rejected. This is a
blemish in Einstein's otherwise perfect special relativity.
\end{abstract}
\textbf{Key words:} special relativity, time contraction, time
dilation, clock paradox, twin paradox, light clock
\newline
\textbf{PACS:} 03.30.+p

\section{Introduction}
As well known, the concept of slowing of a moving clock as
physical interpretation of relative time was first stated in
Einstein's 1905 paper [1] and the lifetime dilation of moving
organisms mentioned in his 1911 paper [2]. This kind of concept
has created the so-called clock paradox or twin paradox. The twin
paradox concerns a pair of twins, one of which aboard a rocket
flies into space for years at a speed fractionally less light
speed and the other remains home. According Einstein's concept,
the travelling twin will be much younger than the other on
returning. This seems to be a paradoxical consequence since motion
is relative. As well known, Einstein $``$solved the paradox by
taking into account the influence on clocks of the gravitational
field, which is relative to (the accelerated system) $K'$.$''$ [3]
He meant that the acceleration and deceleration effects and the
gravity effect are the causes creating an asymmetry of aging of
the twins. The paradox has even led a minute minority of
physicists, for example, Herbert Dingle [4], to argue that
Einstein's relativity must be false. One of the long debated
questions is whether general relativity is required to solve the
paradox. If special relativity is tenable as the overwhelming
majority of physicists believes, what mistake might be ever made
by Einstein?

In search of an answer to the question, the present author thinks
of Einstein's discussion of time measurement. He stated: $``$one
determines the rate of the moving clock by comparing the position
of the hands of this clock with the positions of the hands of
other clocks at rest and if it moves with a velocity approximating
the velocity of light, the hands of the clock would move forward
infinitely slowly.$''$ [2] He asserted also: $``$By definition, to
measure the time interval during which an event takes place means
to count the number of time periods indicated by the clock from
the beginning till the end of the event in question.$''$ [5]
Clearly, his way to measure time is identical with Newton's. It
seems that the concept of the clock that he retained is equivalent
to Newton's concept of absolute time that he rejected. The author
discovers that a geometric illustration of the Lorentz
transformations and a graphic display of similarity between
divisions of a ruler and ticks of a clock can make the matter
clear.

\section{Geometric illustration of the Lorentz transformations}
Equivalently to Minkowski's formulation of special relativity, we
consider two four-dimensional systems $(x,y,x,w)$ and
$(x',y',x',w')$ in relative translational motion and assume that
the origin of the latter moves at the light speed $c$ in
accordance with the following equation:
\begin{equation}
\label{eq1}
c^2t^2=x_0^2+y_0^2+z_0^2+w_0^2, \mbox{ }w_0=c\tau_0
\end{equation}
where $\tau_0$ is the proper time. This equation expresses
mathematically that a free object moves at the maximum speed $c$
at which it can travel in the four dimensions $(x,y,x,w)$ where
the variable $w$ is much like the spatial ones, $x$, $y$ and $z$
[6]. From Eq.(1) we can draw a figure shown as Fig.1 to illustrate
the Lorentz transformations. In this figure only two-dimensional
systems $(x,w)$ and $(x',w')$ are drawn for the purpose of this
article and $x_0=vt$ is given. From the projective relations of
the coordinates in the figure, we can write intuitively with ease
the Lorentz transformations for the points on the $x'$ axis and an
identity:
\begin{figure}[htbp]
\centerline{\includegraphics[width=3.0in,height=2.0in]{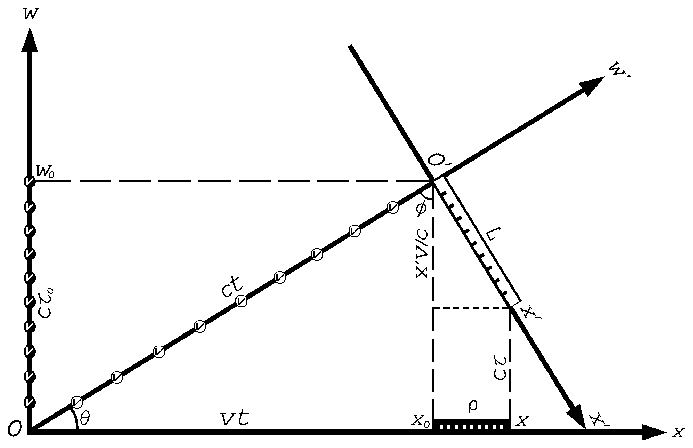}}
\label{Fig.1} \caption{Geometric illustration of the Lorentz
transformations and graphic display of similarity between
divisions of the ruler and ticks of the clock denoted by the
\textit{V} symbols.}
\end{figure}
\begin{equation}
\label{eq2}
x'=\frac{x-vt}{\gamma}, \mbox{ }
\gamma=\cos\phi=\sqrt{1-v^2/c^2}
\end{equation}
\begin{equation}
\label{eq3}
t=\frac{\tau+x'v/c^2}{\gamma}
\end{equation}
\begin{equation}
\label{eq4}
x=\frac{x'+v\tau}{\gamma}
\end{equation}
\begin{equation}
\label{eq5}
\tau=\frac{t-xv/c^2}{\gamma}
\end{equation}
\begin{equation}
\label{eq6}
c^2t^2+x'^2=c^2\tau^2+x^2
\end{equation}
The last equation can be changed into
\begin{equation}
\label{eq7}
c^2t^2-x^2=c^2\tau^2-x'^2
\end{equation}
Thus, in general, including $y$ and $z$, we have
\begin{equation}
\label{eq8}
c^2t^2-x^2-y^2-z^2=c^2\tau^2-x'^2-y'^2-z'^2
\end{equation}
For a photon or a light pulse, we have
\begin{equation}
\label{eq9}
c^2t^2-x^2-y^2-z^2=c^2\tau^2-x'^2-y'^2-z'^2=0
\end{equation}
which describes the principle of constancy of light speed. The
$\tau$ is often written as $t'$ in the literature of relativity.

Let a ruler be located on the $x'$ axis. From Fig.1 we can see
directly the length contraction of the moving ruler:
\begin{equation}
\label{eq10}
\rho=L\gamma=L\cos\phi
\end{equation}
Obviously the length contraction of the ruler is a reciprocal
projection effect between the two equivalent inertial systems in
relative motion and the division number of the ruler is
independent of the relative speed $v$ and hence invariant under
the Lorentz transformations. The length interval between two
neighboring divisions of the moving ruler at the speed $v$ in the
$x$ direction contracts by the factor $\gamma$. Similarly, let a
clock be located at the origin of the moving system and
synchronized with one clock at rest in the stationary system when
the origins of the two systems are coincide at $t=0$. For the
moving clock, in his 1905 paper, Einstein derived from Eq.(3) the
time contraction equation [1]
\begin{equation}
\label{eq11}
\tau_0=t\gamma=t\cos\phi
\end{equation}
which is similar to Eq.(10) in form. This contraction can also be
seen directly in Fig.1. Its inverse process is referred to as
dilation of the time $\tau_0$. According to similarity between
space and time and correspondence between a ruler and a clock, it
is evident that the ticks of the moving clock are nothing but the
correspondences of the divisions of the ruler. Obviously the
length contraction of the ruler is merely attributed to
contraction of the length interval between its two neighboring
divisions, so, similarly, the time contraction of the moving clock
is merely attributed to contraction of the time interval between
its two consecutive ticks and the number of the ticks is also
independent of the speed $v$ and hence invariant under the Lorentz
transformations. But we are unable to employ an existing clock to
record the time $\tau$ which depends on both the speed $v$ and the
location $x'$. Therefore, in order to record it, we need devise an
alternative clock.

\section{Time $\tau$ and $\tau$-clock}
The clocks that we can devise to indicate the time $\tau$ are
shown in Fig.2. In the figure, on the largest circles on the clock
faces, the arc length swept out by the hand
\begin{figure}[htbp]
\centerline{\includegraphics[width=4.6in,height=1.8in]{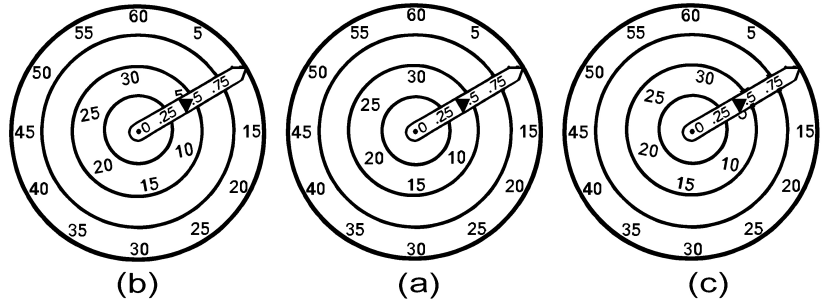}}
\label{Fig.2} \caption{The $\tau$-clocks moving in the $x$
direction, for example, running at $\gamma=0.5$: (a) at the origin
of the moving system, (b) and (c) at the positions a light-second
away from the origin on the negative and the positive $x'$ half-
axises.}
\end{figure}
represents time $t$ in the stationary system. The $\gamma$ values
(0, 0.25, 0.5 and 0.75) on the clock's hand represent the
different relative speeds. The readings in seconds on the circle
to which the sliding arrow points represent the duration of an
event in the moving system judged by an observer in the stationary
system. For example, for the case of relative speed 0.866$c$,
namely, $\gamma=0.5$, Fig.2(a) shows that the clock at the origin
of the moving system ticks 0.5 seconds and Fig.2(b) and Fig.2(c)
show that the clocks a light-second away from the origin tick the
same but run 0.866 seconds later and earlier than the one at the
origin, respectively. The counterclockwise and clockwise angular
displacements of the readings $\tau$ depend on the term $x'v/c^2$
in Eq.(3). Thus this kind of clock has dual functions for
recording both $t$ and $\tau$ simultaneously and hence meets the
relativistic requirement of the equivalence of reciprocal $``$time
measurements$''$ by two different inertial observers. This kind of
clock may be called as $``$$\tau$-clock$''$ to distinguish it from
traditional clocks which record only the time in the stationary
system or systems at very low relative speed. With $\tau$-clocks
both of the twins can think the other's one ticks a less time
period. No paradox! The $\tau$-clock is useless in practice but
essentially important for properly understanding special
relativity.

\section{Behavior of a light clock}
The discussion of the behavior of a moving light clock is also
significant for understanding special relativity. Imagine a light
clock which is composed of a pair of parallel mirrors reflecting
light pulses separated by the distance $d=cT_p/2$ that light
travels in half a second, here $T_p=1$ second. There are a source
of light pulses and a light detector on one mirror for producing
every tick signal. When the vertical light clock moves at uniform
speed $v$ in the horizonal $x$ direction, according to Eq.(9), we
have
\begin{equation}
\label{eq12} c^2(T_p/2)^2-(vT_p/2)^2-y^2(T_p/2)=0
\end{equation}
\begin{equation}
\label{eq13} c^2(\tau_p/2)^2-y'^2(\tau_p/2)=0
\end{equation}
where $y'=y$. Thus, different from the real length $d$, the
$``$relative path$''$ of the pule between the top and bottom
mirrors in the $y$ direction is
\begin{equation}
\label{eq14}
d'=y(T_p/2)=c\gamma T_p/2=c\tau_p/2=\gamma d=d\cos\phi
\end{equation}
that satisfies the equation of the right triangle
\begin{equation}
\label{eq15}
d'^2+v^2=c^2
\end{equation}
here $d'$ is the contracted light path in the $y$ direction judged
by an observer in the stationary system.
\begin{figure}[htbp]
\centerline{\includegraphics[width=4in,height=2in]{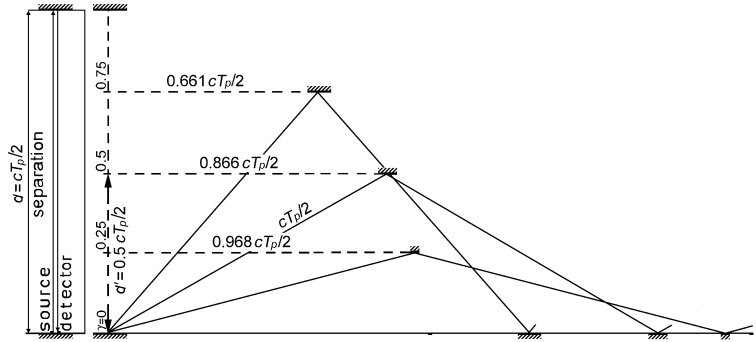}}
\label{Fig.3} \caption{Illustration of the behavior of the light
clock by relative travel paths of the light pulse for different
$\gamma$ values.}
\end{figure}
It is necessary to emphasize again that the time contraction of
the moving light clock is merely attributed to the contraction of
the period between its two consecutive ticks. This means that the
light pulse hits the top mirror when the relative light path
equals $d'$ and the detector produces a tick signal when it gets
to bottom mirror. Fig.3 illustrates the behavior of the light
clock by relative travel paths of the light pulse for different
$\gamma$ values. Thus the explanation of time dilation by the
light clock in textbooks, such as Feynman Lectures on Physics
volume 1 [7], is wrong because of ignoring the relativistic
contraction of the travel path of the light pulse.

\section{Conclusion}
We have seen that, like the division number in a moving ruler, the
tick number of a moving clock is independent of its relative speed
and hence invariant under the Lorentz transformations. So the hand
of the moving clock never runs slow but the time interval between
its two consecutive ticks contracts. Thus it is Einstein's concept
of slowing of the hands of moving clocks to create the clock
paradox or twin paradox. Regrettably, his way to measure time
$\tau$ is identical with Newton's. In other words, the concept of
the clock that he retained is equivalent to Newton's concept of
absolute time that he rejected.

Similar to the clock's ticking, the heart beating of the
travelling twin never runs really slow but according to special
relativity the beating period between two consecutive beats
contracts. Since the human lifetime may be considered as being
basically determined by the total number of the heart beats, no
one of the twins will become younger than the other when they
reunite. Thus, the general relativity is not required to resolve
the twin paradox. It is Einstein's assertion of slowing of the
hands of moving clocks to create the paradox and hence make many
physicists believe slowing of all physical processes with the
increased speed, including chemistry reactions, nuclear reactions,
life process and others. This is a blemish in Einstein's otherwise
perfect special relativity.

Finally, the author would say that if the present conclusion is
correct, it will prohibits us from believing the assertion that
some related experiments [8,9] have confirmed the Einstein's
prediction of slowing of moving clocks that creates the clock
paradox.

\end{document}